\documentclass{sig-alternate}
\usepackage{epsfig}
\begin{document}
\conferenceinfo{ICTKS}{'08 Huntsville, AL}
\title{Toward Understanding Friendship in Online Social Networks}
\numberofauthors{2}
\author{%
\alignauthor Dmitry Zinoviev\\
       \affaddr{Mathematics and Computer Science Department}\\
       \affaddr{Suffolk University}\\
       \affaddr{Boston, Massachusetts 02421, USA}\\
       \email{dmitry@mcs.suffolk.edu}%
\alignauthor Vy Duong\\
       \affaddr{Mathematics and Computer Science Department}\\
       \affaddr{Suffolk University}\\
       \affaddr{Boston, Massachusetts 02421, USA}\\
       \email{duo11801@suffolk.edu}%
}
\date{25 February 2009}
\maketitle

\section*{ABSTRACT}
All major on-line social networks, such as MySpace, Facebook, LiveJournal, and Orkut, are built around the concept of friendship. It is not uncommon for a social network participant to have over 100 friends. A natural question arises: are they all real friends of hers, or does she mean something different when she calls them ``friends?'' Speaking in other words, what is the relationship between off-line (real, traditional) friendship and its on-line (virtual) namesake? In this paper, we use sociological data to suggest that there is a significant difference between the concepts of virtual and real friendships. We further investigate the structure of on-line friendship and observe that it follows the Pareto (or double Pareto) distribution and is subject to age stratification but not to gender segregation. We introduce the concept of digital personality that quantifies the willingness of a social network participant to engage in virtual friendships.
\vskip\baselineskip\noindent
{\bf Keywords:} Online Social Network, Friend, Overlay, Demographics.

\section{INTRODUCTION}
Massive online social networks (MOSN) have become increasingly popular in the last several years. The largest online social networks, such as Facebook, LinkedIn, and MySpace, have tens and hundreds of millions of member accounts (some of which, however, can be dormant). Direct descendents of the Internet forums and chatrooms of the 1980s and 1990s, they serve as venues for communication, socialization, cooperation, learning, electronic commerce, and even cybercrime.

All modern MOSNs have similar internal structure. Each MOSN can be represented as a mesh of interconnected nodes, where every node represents a member account, and the connections between the nodes represent relationships between the members. A member account typically represents one human user, but it is not uncommon for a single user to create multiple online personalities (e.g., for the purpose of separating business and leasure) or for a group of users (say, working together on a project or representing a closely connected social group) to create a single member account. For the rest of the paper we assume that each member account corresponds to one human user.

A connection between two MOSN members is typically established if and when the two believe to share something in common, such as a common background, common friends, common interests and activities---or are interested in exploring friends, interests, and activities of others. A connection can be implicit---through online communities (member groups by interests and causes~\cite{spertus2005,watts2002})---or explicit. Explicitly, a MOSN member can have one or more so-called ``friends' lists,'' to which the connections are added according to certain rules. It is mainly through the online (``virtual'') friendship that the communications between the network members occur (which does not exclude the possibility of peer-to-peer communications between the members who are not formally virtual ``friends'').

In this paper, we will show that there is a significant difference between the concepts of virtual and real friendships. We will present quantitative data from a real MOSN that demonstrate that not all ``friends'' on the Internet are real friends and that the overlap between the circles of offline friends and online friends exists but is not overwhelming. We will analyse the demographics of the online friendship.

In addition, we introduce the concept of quantitative personality that measures an MOSN user's willingness to engage in virtual friendships, and discuss its relevance to the traditional offline friendship. 

\section{RELATED WORK}
The significant difference between virtual and real friendship as such has been explored in~\cite{boyd2006}, however, no quantitative evidence was presented.

The mechanisms of ``friendship'' allow the users to establish new connections and enable social searching (locating and maintaining offline connections online, in a virtual setting, to learn more about them, date with them, or even engage in casual sex)~\cite{lampe2006}. It has been shown that through ``friendships'' users impact their friends' decisions~\cite{brzozowski2008}, affect the predictability of the friends' actions and adopt behaviours exhibited by their friends~\cite{crandall2008}, and influence the behavior patterns of their friends~\cite{maia2008,singla2008}.

Some numerical data that describe the composition of the online friends' body and the relationship between the online and offline friendship have been presented in~\cite{gilbert2008, golder2007,ploderer2008}. The insignificance of the online/offline overlap is mentioned in~\cite{ploderer2008}. Golder et al.~\cite{golder2007} study the distribution of number of friends per user in Facebook and explain that ``people add friend links for a variety of reasons, not always for reasons that imply the pair are friends in the conventional sense.'' Finally, Gilbert et al. ~\cite{gilbert2008} explore the gender aspect of online friendship and separate online friendships into ``strong ties'' and ``weak ties.'' However, the separation is based on the friend-to-friend message statistics rather than on the social and historical origins of the friendship links.

\section{FRIENDSHIP IN ONLINE SOCIAL NETWORKS}

While initially the meaning of a ``friendship'' relationship in MOSNs was to represent a personal friendship (``affection arising from mutual esteem and good will''~\cite{webster1913}), the simplicity of befriending people online soon distorted the original intention. Being a friend in a MOSN can be considered a necessary but not a sufficient condition for being a true online friend~\cite{golder2007}. To distinguish between true friends (online of offline) and registered online ``friends,'' we will refer to the latter as contacts.
\begin{figure}[tb!]\centering
\epsfig{file=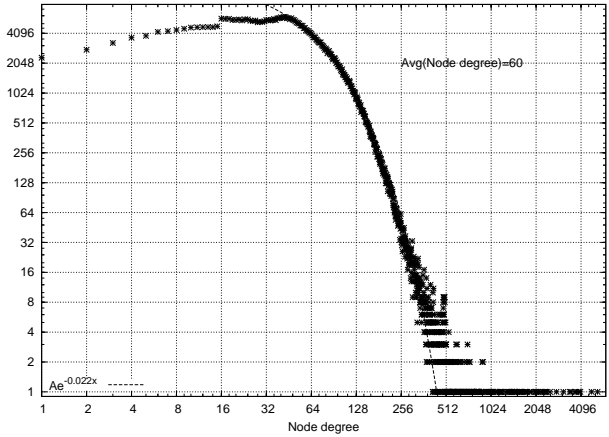,width=\columnwidth}
\caption{\label{friends}Distribution of number of friends (users with X friends against the number of friends) in the subset A}
\end{figure}

The reasons that stimulate members to engage in online ``friendships'' range from real sincere friendship to the inability to say ``no'' and the desire to explode popularity~\cite{boyd2006}.  In many online social networks (e.g., LiveJournal) ``friendship'' is not even reciprocal, that is, there is no need to obtain someone's permission to declare her as a ``friend.'' As a result, it is not uncommon for up to 15\% of MOSN members to have more than 100 contacts (to have the degree of over 100, Figure 1). We claim that most of these contacts are not friends in the conventional sense. In fact, all relationships can be roughly separated into three groups: true friends (strong tie, according to~\cite{gilbert2008}), good acquaintances (medium strength ties), and random acquaintances (weak ties).

To validate our hypothesis, we used two subsets (groups) of members of Odnoklassniki.Ru (Classmates)---the leading Russian-language MOSN. The group A contains ca. 500 thousand members, or ~5\% of the total MOSN population (at the time of writing), together with all the ``friendship'' connections among the members and the basic demographics (age and gender). 

Note that group A has an unusual---geometric---node degree distribution (Figure~\ref{friends}). More often, major social networks follow the Pareto or double Pareto power law~\cite{novak06,reed01}.

The other group B (the subset of A) contains 150 members with additional information about the nature of their ``friendships.'' This additional information was obtained by interviewing ca. 1000 network members, of which only a small fraction (15\%) responded. Each member in the group B was asked the following four questions (in Russian):

\begin{enumerate}
\item How many real friends or relatives do you have on your contact list?
\item How many of the remaining members on the list are good acquaintances of yours?
\item How many of the remaining members on the list are random acquaintances of yours? 
\item How many of your real friends do not have an account at Odnoklassniki.Ru or are not on your contact list?
\end{enumerate}

The purpose of the fourth question was to estimate the fraction of the overall true friends' body that is also on the member's MOSN contact list. It was left up to the respondents to decide which contacts belonged to what group.

\section{ONLINE FRIENDSHIP AS EXTENDED FRIENDSHIP}

\begin{figure}[tb!]\centering
\epsfig{file=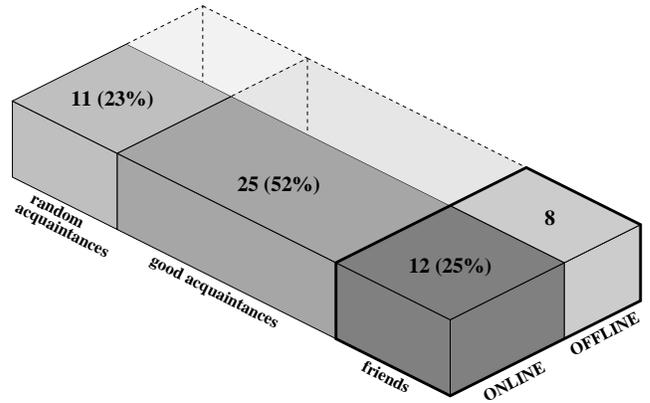,width=\columnwidth}
\caption{\label{friendship}Structure of the online ``friendship'' relationship: average numbers and fractions of real friends (both online and offline), good acquaintances, and random acquaintances
}
\end{figure}

The results of the poll are shown in Figure~\ref{friendship}. The relative sizes of the blocks are proportional to the average numbers of MOSN members in the corresponding classes.

An average group B member has 48 contacts (vs. 60 contacts on the list of a group A member). Of them, only $F_{on} = 12$ (or 25\%) are recognized as real friends. More than 50\% of contacts are good acquaintances ($F_{ga} = 25$). The remaining $F_{ra} = 11$ contacts are considered random acquaintances. There is an exceptionally low correlation between the total number of contacts F and Fga and low negative and positive correlations between F and Fon and F and Fra, respectfully. This means that the fraction of good acquaintances does not depend on the size of the ``first circle'' of contacts, while the fraction of good friends slowly diminishes at the expense of the gradually growing fraction of random acquaintances. The observed behavior is a variation of the ``rich get richer'' scheme, where random acquaintances---the ``true friendship' ballast''---are the equivalence of wealth.

Another interesting observation is that an average group B member has a total of 20 real friends. Of them, 12 friends (60\%) are on the MOSN contact list and eight are not, i.e., the online ``friendship'' somewhat absorbs traditional friendship (and also substantially extends it). The fraction of ``offline'' friends has a substantial negative logarithmic correlation $r = -0.3$ with the number of contacts suggesting that the MOSN members who are socially active online either pull their offline friends into the MOSN or have fewer offline friends in the first place.

We do not have data to compare the online and offline populations of good and random acquaintances. 

\section{AGE STRATIFICATION}

The demographics of the online friendship in the group A is subject to strong age stratification: the members of a certain age prefer to contact members of approximately the same age. If there were no age preferences among the members, then we would expect about 10 times fewer same-age relationships and 10 times more relationships with larger age difference than we observe in the group A (Figure~\ref{agedifference}).

Figure~\ref{agestratification} shows the distribution of group A contacts' age (the maxima, or ``ridges,'' are at the thick black lines). According to the chart, all group A members can be roughly separated into four age categories:

\begin{figure}[tb!]\centering
\epsfig{file=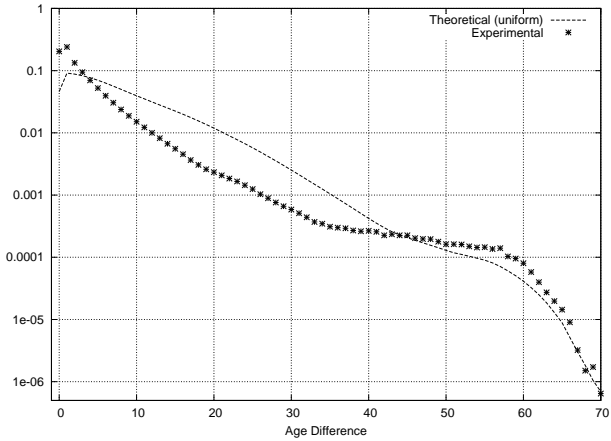,width=\columnwidth}
\caption{\label{agedifference}Experimental and theoretical age difference distribution in all friendships among the group A members (fraction of the number of pairs with age difference X against the age difference)}
\end{figure}

\begin{enumerate}
\item Ages 11--14 mainly befriend peers (``Ridge 1'') and members in the range 19--30 year (teenage drive, ``Ridge 3'').
\item Ages 15--41 mainly befriend peers (classmates, schoolmates, comrades-in-arms, ``Ridge 1''), as well as members in the lower teens (``Ridge 2'') and in the lower 60s (``Ridge 5'').
\item Ages 42--52 mainly befriend peers (``Ridge 1'').
\item Ages 53 and older mainly befriend members in the 20s (children, grandchildren, ``Ridge 4'').
\end{enumerate}

The ``Plateaus'' surrounding the main diagonal in the range from 25 to 55 years probably represent co-workers in their productive ages.

\begin{figure}[tb!]\centering
\epsfig{file=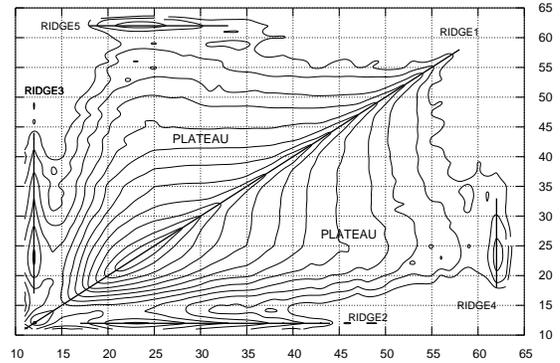,width=\columnwidth}
\caption{\label{agestratification}Age stratification in group A: ages of a X's contacts against X's own age; local maxima are at thick black lines}
\end{figure}

We were not able to detect gender-based segregation within group A: 48\% of the relationships are intergenderal, 29\% are between females, and 23\% are between males.

\section{QUANTITATIVE PERSONALITY}

To quantify the relative importance of a network member (say, ``Alice'') among her immediate contacts, we introduced a quantitative personality $\Pi$---a measure of social popularity/marginality~\cite{zinoviev2008}. Let $\varepsilon$ be all of Alice's immediate contacts (her ``first circle''). Then her quantitative personality $\Pi$ is the logarithm of the ratio of the average number of her contacts' contacts $\rho(\varepsilon)$ to the number of her own contacts $\rho(A)$. If Mi is Alice's contact and $\rho(M_i)$ is the number of $M_i$'s contacts, then:

$$
\Pi = \log [\Sigma \rho(M_i) / \rho(A)] / \rho(A) = \log \Sigma \rho(M_i) - 2\log \rho(A). 
$$

The logarithmic function is used to restrict the range of the ratio that otherwise could be very broad. 

The positive values of $\Pi$ mean that $\rho(\varepsilon) > \rho(A)$, i.e., Alice's  contacts have on average more contacts than Alice herself, and Alice is socially marginal. Conversely, if $\Pi< 0$ then Alice has more contacts than her neighbors and she is socially popular. $\Pi = 0$ is the case of equilibrium when Alice is an average MOSN member. It has been shown in~\cite{zinoviev2008} that the marginals outnumber the popular members approximately by the factor of 10. 

We compared the values of $\Pi$ calculated through the analysis of the MOSN, with our experimental data to determine if this online parameter is relevant to the nature of the online friendship relation and to the members' offline behavior. What we discovered was a reasonably strong negative linear correlation between $\Pi$ and the number of online friends $F_{on}$ ($r = -0.42$), good acquaintances $F_{ga}$ ($r = -0.5$), and random acquaintances $F_{ra}$ ($r = -0.42$). This is consistent with the definition of $\Pi$: smaller $\Pi$ means more contacts.

\setlength{\textheight}{18.5cm}

However, the correlation between $\Pi$ and the number of offline friends Foff is close to zero: the quantitative personality is probably not related to offline friendship.

\section{CONCLUSION}
The concept of friendship is fundamental to all major online social networks.  However, ridiculously large numbers of online friends that we observe in a typical massive online social network (MOSN) make us question the validity of the term ``friendship.'' By interviewing a randomly chosen group of members of Odnoklassniki.Ru (a major Russian social network), we demonstrate that online friendship is an umbrella name for real friendship and various degrees of acquaintanceship and that on average only ~25\% of contacts are recognized as real friends. More popular MOSN members tend to have fewer real offline-only friends (either because they drag their conventional friends into the MOSN or because they had inherently fewer offline friends).

We observe strong age stratification in online ``friendship'': all MOSN members roughly fall into four age ranges with range-specific age preferences; the members of the two largest of the ranges, 15--52 years, prefer to befriend people of about the same age.

We apply the concept of quantitative personality proposed in an earlier paper to the members of Odnoklassniki.Ru, and discover that it is not correlated with the number of real offline friends, which probably means that it is an online-only attribute of social behavior.

\section{ACKNOWLEDGMENTS}
This research has been supported in part by the College of Arts and Sciences, Suffolk University, through an undergraduate research assistantship grant.  The authors are thankful to the members of the Odnoklassniki.Ru online social network for their cooperation.

\bibliographystyle{abbrv}

\bibliography{cs}

\begin{thebibliography}{10}

\bibitem{boyd2006}
D.~Boyd.
\newblock Friends, friendsters, and top 8: Writing community into being on
  social network sites, dec 2006.

\bibitem{brzozowski2008}
M.~Brzozowski, T.~Hogg, and G.~Szabo.
\newblock Friends and foes: Ideological social networking.
\newblock In {\em CHI '08: Proceeding of the twenty-sixth annual SIGCHI
  conference on Human factors in computing systems}, pages 817--820, Florence,
  Italy, 2008. ACM.

\bibitem{crandall2008}
D.~Crandall, D.~Cosley, D.~Huttenlocher, J.~Kleinberg, and S.~Suri.
\newblock Feedback effects between similarity and social influence in online
  communities.
\newblock In {\em Proc. KDD'08}, pages 160--168, Las Vegas, NV, Aug. 2008. ACM.

\bibitem{gilbert2008}
E.~Gilbert, K.~Karahalios, and C.~Sandvig.
\newblock The network in the garden: an empirical analysis of social media in
  rural life.
\newblock In {\em CHI '08: Proceeding of the twenty-sixth annual SIGCHI
  conference on Human factors in computing systems}, pages 1603--1612,
  Florence, Italy, 2008. ACM.

\bibitem{golder2007}
S.~Golder, D.~Wilkinson, and B.~Huberman.
\newblock Rhythms of social interaction: Messaging within a massive online
  network.
\newblock In {\em Proc. 3rd International Conference on Communities and
  Technologies (CT2007)}, East Lansing, MI, June 2007.

\bibitem{novak06}
R.~Kumar, J.~Novak, and A.~Tomkins.
\newblock Structure and evolution of online social networks.
\newblock In {\em Proc. KDD'06}, pages 611--617, Philadelphia, PA, Aug. 2006.

\bibitem{lampe2006}
C.~Lampe, N.~Ellison, and C.~Steinfield.
\newblock A face(book) in the crowd: Social searching vs. social browsing.
\newblock In {\em CSCW '06: Proceedings of the 2006 20th anniversary conference
  on Computer supported cooperative work}, pages 167--170, Banff, Alberta,
  Canada, 2006. ACM.

\bibitem{maia2008}
M.~Maia, J.~Almeida, and V.~Almeida.
\newblock Identifying user behavior in online social networks.
\newblock In {\em SocialNets '08: Proceedings of the 1st workshop on Social
  network systems}, pages 1--6, Glasgow, Scotland, 2008. ACM.

\bibitem{ploderer2008}
B.~Ploderer, S.~Howard, and P.~Thomas.
\newblock Being online, living offline: the influence of social ties over the
  appropriation of social network sites.
\newblock In {\em Proc. ACM 2008 conference on Computer supported cooperative
  work}, pages 333--342, San Diego, CA, 2008.

\bibitem{reed01}
W.~Reed.
\newblock The {P}areto, {Z}ipf and other power laws.
\newblock {\em Economics Letters}, (74):15--19, 2001.

\bibitem{singla2008}
P.~Singla and M.~Richardson.
\newblock Yes, there is a correlation:---from social networks to personal
  behavior on the web.
\newblock In {\em WWW '08: Proceeding of the 17th international conference on
  World Wide Web}, pages 655--664, Beijing, China, 2008. ACM.

\bibitem{spertus2005}
E.~Spertus, M.~Sahami, and O.~Buyukkokten.
\newblock Evaluating similarity measures: a large-scale study in the {O}rkut
  social network.
\newblock In {\em Proc. KDD'05}, pages 678--684, Chicago, IL, Aug. 2005. ACM.

\bibitem{watts2002}
D.~Watts, P.~Dodds, and M.~Newman.
\newblock Identity and search in social networks.
\newblock {\em Science}, 296:1302--1305, May 2002.

\bibitem{webster1913}
N.~Webster.
\newblock {\em Webster's Revised Unabridged Dictionary}.
\newblock Merriam-Webster, Springfield, MA, 1913.

\bibitem{zinoviev2008}
D.~Zinoviev.
\newblock Topology and geometry of online social networks.
\newblock In {\em Proc. of The 12th World Multi-Conference on Systemics,
  Cybernetics and Informatics}, volume~VI, pages 138--143, Orlando, FL, jul
  2008.

\end{thebibliography}

\balancecolumns

\end{document}